  \documentclass[multi]{cambridge6Atight}
  \usepackage{natbib}

  \usepackage{rotating}
  \usepackage{floatpag}
  \rotfloatpagestyle{empty}

\usepackage{amsmath}
  \usepackage{amsthm}
  \usepackage{graphicx}
  \usepackage{mathptmx}
  \usepackage{color}

  \usepackage{makeidx}
  \makeindex

\copyrightline{This material has been published in \textit{The Impact of Binaries on Stellar Evolution}, Beccari G. \& Boffin H.M.J. (Eds.).
This version is free to view and download for personal use only. Not for re-distribution, re-sale or use in derivative works. \copyright\ 2018 Cambridge University Press.}

\begin{document}


  \alphafootnotes
   \author[P Kroupa and T Jerabkova]
     {Pavel Kroupa
    and Tereza Jerabkova}
  \chapter{The Impact of Binaries 
on the Stellar Initial Mass Function}


  \contributor{Pavel Kroupa
    \affiliation{Helmhotz-Institut f\"ur Strahlen und Kern-Physik,
      Universit\"at Bonn and Astronomical Institute, Charles
      University in Prague}
}

  \contributor{Tereza Jerabkova
    \affiliation{European Souther Observatory,
    Karl-Schwarzschild-str. 2, 85748 Garching, Germany}}

 \begin{abstract}
   The stellar initial mass function (IMF) can be conveniently
   represented as a canonical two-part power law function and is
   largely invariant for star formation regions evident in the Local
   Group of galaxies. The IMF is a ``hilfskonstrukt''. It is a
   mathematical formulation of an idealised population of stars formed
   together in one star formation event, that is in an embedded
   cluster. The nature of the IMF -- is it a {\it probability density}
   or an {\it optimal sampling} distribution function -- is raised.
   Binary stars, if unresolved, have a very significant influence on
   the counts of low mass stars and need to be corrected
   for. Especially important is to take care of the changing binary
   fraction as a result of stellar-dynamical evolution of the embedded
   clusters which spawn the field populations of galaxies, given that
   the binary fraction at birth is very high and independent of
   primary-star mass. At the high mass end, the shape of the IMF is
   not much affected by unresolved companions, but the high
   multiplicity fraction amongst massive stars leads to a substantial
   fraction of these being ejected out of their birth clusters and to
   massive stars merging. This explains the top-lightness of the IMF
   in star clusters in M31.  In close binaries also the masses of the
   components can be changed due to mass transfer.  Correcting the
   star counts in R136 in the 30~Dor region for ejected stars, and a
   stellar-dynamical analysis of globular clusters and of ultra
   compact dwarf galaxies, together with the most recent star-counts
   in low-metallcity young massive clusters, all together imply the
   IMF to become top-heavy with decreasing metallicity and above a
   star-formation-rate density of about $0.1\,M_\odot/($yr pc$^3)$ of
   the cluster-forming cloud core. This is also indicated through the
   observed supernova rates in star-bursting galaxies. At the same
   time, the IMF may be bottom light at low metallicity and
   bottom-heavy at high metallicity, possibly accounting for the
   results on elliptical galaxies and ultra-faint dwarf galaxies,
   respectively.
 \end{abstract}

\section{Introduction}
\label{sec:KJ_sec1}

If the number of stars in the initial-stellar-mass
interval $m,m+{\rm d}m$ is ${\rm d}N$ then
\begin{equation}
{\rm d}N=\xi(m)\,{\rm d}m,
\end{equation}
where $\xi(m)$ is the stellar initial mass function (IMF), which is,
formally, the complete ensemble of stars with their initial masses
which form together.  Because stars do not form at the same instant in
time, the stellar initial mass function (IMF) does not exist as
such. Indeed, even if we were able to measure the mass of each star,
it is not possible to empirically measure the IMF: at any given time a
stellar population is observed, some massive stars have either already
evolved or been ejected from the system, the low-mass stars may not
have formed yet in entirety and/or some have been lost through stellar
dynamical processes, and stellar mergers may have occurred.

The IMF is a mathematical short-cut, a '{\it hilfskonstrukt}',
allowing the modelling of stellar populations, the stars of which form
in spatially and temporarily correlated star formation events (CSFEs),
i.e. in density maxima in molecular clouds which are fed by convergent
filamentary gas flows \citep{Kirk+13, Hacar+17b, Burkert17}. Theory
has not been very successful in quantifying how the gas in a galaxy
transforms into stars, due to the immense complexities and multitude
of physical processes involved, but observations inform that stars
form in CFSEs, which for practical purposes can be called embedded
clusters \citep{Lada2003a, Lada10, Gieles+12, Megeath16,
  Kroupa+18}. These are typically of sub-pc scale and assemble over a
time-scale of a Myr \citep{Marks12b}, having spatial dimensions
comparable to the molecular cloud filaments and the intersection
thereof \citep{Andre+16, Lu+18}. These embedded clusters appear to
form mass-segregated as ALMA and other observations have shown
\citep{KM11, Plunkett+18,AR18} and may contain from a dozen binaries
to many millions of stars, as a result of the large range of
gravity-assisted density fluctuations, filaments and inflows in
molecular clouds, which themselves are merely momentary condensations
of the complex interstellar-medium of a galaxy.  Data on globular
clusters also indicate these to have formed mass-segregated
\citep{Haghi15,Zonoozi2017}.  The flat birth radius--mass relation
\citep{Marks12b} and primordial mass segregation may be directly
related to the existence of the most-massive-star vs
embedded-cluster-stellar-mass (the $m_{\rm max}-M_{\rm ecl}$) relation
\citep{weidner2006a}.

Star-formation in the embedded clusters is feedback regulated, most
likely as a result of the sum of outflows and stellar radiation
compensating the depth of the potential of the embedded cluster and
individual proto-stars.  This is evident in the majority of the gas
being expelled as shown by observations \citep{Megeath16} and
magneto-hydrodynamical simulations \citep{MM12, Bate+14, Federrath+14,
  Federrath15, Federrath16}. Noteworthy is that observed CSFEs
spanning the range of a few~10 to a few $10^5\,M_\odot$ in stars (the
Taurus-Auriga embedded clusters, the Orion Nebula Cluster, NGC3603 and
R136, all of which have been observed in much detail) are dynamically
and physically well reproduced if the star-formation efficiency (SFE)
is about 33~per cent, the gas expulsion occurs with about 10~km/s
\citep{Whitmore+99, Zhang+01, Kroupa05, Smith+05} and the embedded
phase lasts about 0.6~Myr \citep{kroupa2003a, KAH, BK13, BK14, BK15b,
  BK17}. It is an interesting problem to understand which mechanism
disperses the gas from these systems.  Observations suggest that even
T~Tauri associations loose their residual gas on a time scale of about
a Myr \citep{Neuhaueser+98}, and \cite{Hansen+12} perform
gravo-radiation-hydrodynamics simulations of low-mass star formation
in turbulent clouds according to which about 2/3rd of the gas is blown
out of the low-mass embedded clusters. The individual embedded
clusters therefore should expand significantly within a few~Myr
\citep{BK17}.  Direct evidence for this expansion has been
documented. \cite{Getman+18} note that the core radii of a large
ensemble of very young clusters expand by a factor of about 10 within
1~to~3.5~Myr and and previously \cite{Brandner08} provided a useful
collation of the size changes for very young clusters as a function of
their age. \cite{BK17} show that the expansion can be achieved only
via the expulsion of residual gas if SFE$\;\approx 1/3$ because
binary-star heating and mass loss through evolving stars do not
suffice, even in unison, to expand the clusters sufficiently within
the short time ($\approx 10\,$Myr). The gas expulsion unbinds a large
fraction of modest-mass embedded clusters \citep{Brinkmann17} for
which observational evidence has been found \citep{Weidner+07} with
potentially important implications for the structure of galactic disks
\citep{Kroupa02b}.  This leads to an understanding of OB associations
as forming from many expanding post-gas-expulsion embedded clusters,
all of which formed over a few Myr within a molecular cloud which may
itself be expanding or contracting such that the kinematical field of
the young stars is complex and may not show an overall expansion of
the OB association (e.g. \citealt{WrightMamajek18}).

Since the time scales of stellar and galaxy evolution, which much of
astrophysics is concerned with, are longer than a few Myr, the stellar
populations may, to a good degree of approximation, be described by
the IMF, if, for example the evolution of a star cluster or its
spectral energy distribution is to be quantified.  Because the
appearance and evolution of astrophysical systems sensitively depends
on the relative fraction of the short-lived massive versus the
long-lived low-mass stars, much emphasis has been placed on
constraining the detailed shape of the IMF and of the possible
variation of this shape with physical conditions of the star-forming
gas. Reviews of this topic can be found in \cite{Chabrier03, Bastian10,
  Kroupa13}.

\section{The nature of the IMF of stars and in galaxies}
\label{sec:KJ_natIMF}

Here an overview of the IMF is given in view of recent
developments. Before continuing, the important point needs to be
stressed that the IMF (normalised to e.g. unit stellar mass) is
mathematically strictly identical to the galaxy-wide IMF (gwIMF, the
IMF of all stars forming in a galaxy and normalised in the same
manner) if the process of star formation is mathematically equivalent
to random sampling from an invariant IMF without constraints.  Only in
this case will many low-mass CSFEs together have the same IMF as one
massive embedded cluster with the same number of stars.  That is,
\begin{equation}
\mathrm{gwIMF} = \mathrm{IMF}\,,
\label{eq:TJ_gwIMF}
\end{equation}
if the invariant IMF is a probability density distribution function
(PDF), where invariant means, 
\begin{equation}
  \mathrm{IMF}(x_1,\varrho_1, t_1, T_1, M_{\mathrm{tot,1}},\dots) = 
\mathrm{IMF}(x_2,\varrho_2, t_2, T_2, M_{\mathrm{tot,2}},\dots)\,,
\label{eq:invIMF}
\end{equation}
for any $\mathrm{parameter}_{1,2}$.  That is, the IMF, which is
invariant in any local physical parameter (in one CSFE, i.e. in one
embedded cluster), will result in the same gwIMF. Note that this
implicitly assumes that massive stars can form in very low density
regions. 

Thus, by constraining observationally gwIMF, this one fundamentally
important concept is tested \citep{Kroupa13}. Indeed, the
Milky-Way-field IMF, extracted from the Galactic field stellar
population, has been found to be top-light relative to the canonical
IMF (Fig.~\ref{fig:KJ_varIMF}), as shown by \cite{Scalo86} and
\cite{RybizkiJust15}. For dwarf late-type galaxies with small
star-formation rates (SFRs), it appears to be top-light (lacking
massive stars,\citealt{Pflamm09b, Lee+09, Watts+18}), and for
late-type galaxies with high SFRs and elliptical galaxies wich formed
as major early star-bursts, the galaxy-wide IMF is top-heavy
\citep{Matteucci94, GM97, vanDokkum08, Gun11, Romano17}. The survey by
\cite{Hsu+12, Hsu+13} of very young stars in the the low-density
star-forming region of the Orion A cloud south of the dense Orion
Nebula Cluster, where only low-mass embedded clusters are found, has
revealed significant evidence for a top-light IMF which differs from
the IMF found in the Orion Nebula Cluster.

These results suggest that the IMF may not be a probability density
distribution function. Progress on this issue has been
achieved recently by interpreting the IMF as an optimal sampling
distribution function \citep{Kroupa13, Schulz15, YJK17}. {\it A variable
$x$ is distributed according to an optimally
sampled distribution function if, upon binning and independent of bin
size, the Poisson scatter is zero in each bin.}

The physical interpretation of optimal sampling being perfect feedback
self-regulation of the star-formation process possibly with a high
regularity of stellar masses along their hosting filaments and
fibres. While optimal sampling is an extreme mathematical assumption,
and nature most likely has some randomisation process when embedded
clusters form, the lack of variation of the shape of the stellar IMF
and the lack of scatter in the $m_{\rm max}-M_{\rm ecl}$ relation
indicate that star-formation may be closer to optimal than random
sampling \citep{Kroupa13}.

This is intimately linked to the existence of the
$m_{\rm max}-M_{\rm ecl}$ relation, according to which an embedded
cluster of stellar mass $M_{\rm ecl}$ has a most massive star of mass
$m_{\rm max}$ \citep{Weidner13b, Ramirez16, Megeath16, Stephens+17}
that can be potentially physically related to the density of fiber
structures in molecular clouds \citep{Hacar+17, Hacar+18}.  Because of
this relation, Eq.~\ref{eq:TJ_gwIMF} does not hold, because many
low-mass CSFEs add only low-mass stars while the few if any massive
CFEs do not add enough massive stars to the galaxy if the CFSEs are
distributed as a Salpeter-like power-law embedded cluster mass
function \citep{YJK17}.  The existence of the
$m_{\mathrm{max}}-M_{\mathrm{ecl}}$ relation may also pose important
implications for the existence of multiple populations/star-bursts in
star clusters, as discussed by \cite{BJK17} and \cite{Kroupa+18}.

\section{The stellar mass--luminosity relation} 
\label{sec:KJ_MLrel}

Stellar masses cannot be measured directly in most instances, and so
the IMF can only be inferred by transforming the luminosity of a star
to its mass using the mass--luminosity relation (other techniques such
as measuring the gravity at the relevant stellar photosphere are
unrealistic for large ensembles of stars). Given the uncertainties
involved, this only works reliably for main-sequence
stars. Immediately the problem becomes apparent that low-mass stars
have not yet reached the main sequence when the massive stars have
already evolved away from it, which is why $\xi(m)$ is but a
hilfskonstrukt.  For an ensemble of single main-sequence stars, which
can be created for an observed population by artificially correcting
all stars to their zero-age main sequence, we have
\begin{equation}
  \xi(m) = {{\rm d}N \over {\rm d}m} = -{{\rm d}N \over {\rm d}M_x} \,
  {{\rm d}M_x \over {\rm d}m},
\label{eq:KJ_xi}
\end{equation}
where $M_x$ is the absolute magnitude of a star of mass $m$ in the
photometric pass- (e.g. V-) band. The $M_x(m)$ function is
theoretically uncertain, in particular near $m=0.3\,M_\odot$, below
which the stars become fully convective and molecular hydrogen becomes
an important opacity source and contributor to the mean molecular
weight. The theory of the internal constitution of stars remains
uncertain due to the complex physics of convection, radiation
transport, rotation and magnetic fields. For $m<0.1\,M_\odot$ the
cores of the stars become electron degenerate, causing the $M_x(m)$
function to become very steep, because a small change in mass does not
lead to a corresponding change in the central density. Thus, small
changes in $m$ cause large changes in $M_x$, with major
uncertainties. Empirical constraints on the $M_x(m)$ function, in
connection with theoretical models, have yielded good constraints on
the IMF \citep{KTG93, Kroupa02, Chabrier03}.  A noteworthy outcome of
this work was the realisation that the function $-{\rm d}m/{\rm d}M_x$
has a pronounced maximum near $M_V\approx11.5 (m\approx 0.3\,M_\odot)$
(\citealt{KTG90}, the KTG peak). It is evident in the stellar
luminosity function, $\Psi_V={\rm d}N/{\rm d}M_V$, of all resolved
stellar populations \citep{Kroupa02, KT97}.  The implication is that
the IMF does not have a turn-over at this mass. It is noteworthy that
just by simply counting stars on the sky to construct $\Psi_X$ their
internal constitution is probed.

\section{Binary stars} 
\label{sec:KJ_bins}

\begin{figure}
    \includegraphics[scale=0.3]{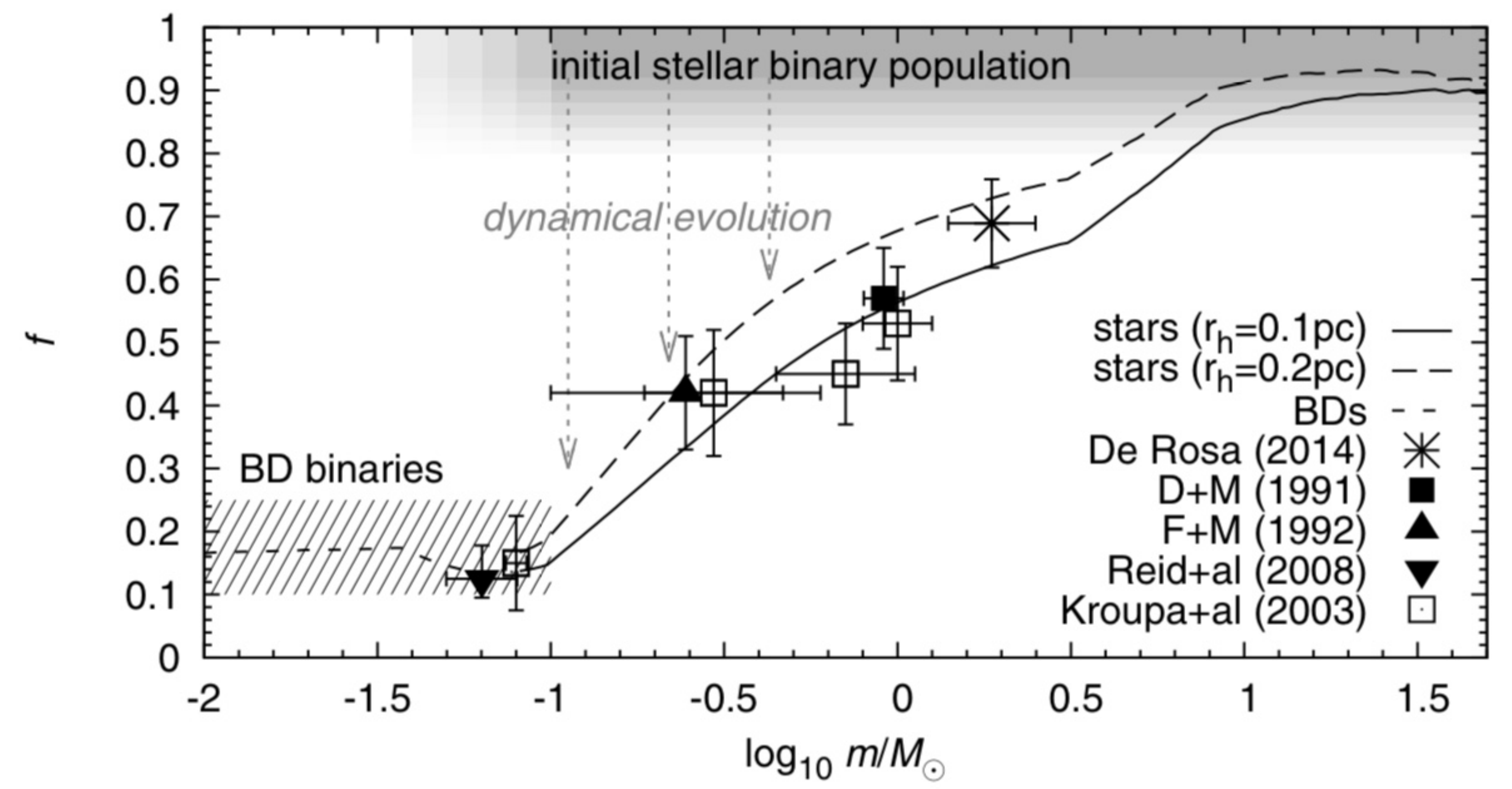}
    \caption[The binary fraction as a function of primary star mass]
    {The binary fraction, $f$, as a function of primary star mass,
      $m$. The data (shown by various symbols) demonstrate that for
      old, main sequence Galactic field stars the binary fraction
      decreases with decreasing primary mass. Young, i.e. TTauri
      stars, however have a binary fraction near unity independently
      of mass. This is shown as the shaded region termed ``initial
      stellar binary population''. It extends to equal-age massive
      stars which have a very high multiplicity fraction as well
      \citep{Sana+12,MDiS17}. The difference for young stars with
      $f\approx1$ and for brown dwarfs (BDs) with $f\approx 0.15$
      implies that BDs form with stars but follow their own
      distribution functions \citep{Marks+15}. It is definitely
      incorrect to argue that the trend of decreasing $f$ with
      decreasing $m$ implies that BDs form as an extension of the
      stellar population. An observational approach to test this
      statement is suggested by \cite{Marks+17}.  The old main
      sequence field stellar binary properties can be well understood
      if all stars form as binaries in embedded clusters with
      half-mass radii $r_{\rm h}$ \citep{Marks12b, Belloni+18} in
      which the binary systems are dissolved over time as the clusters
      evolve (``dynamical evolution'' arrows).  The data are, from top
      to bottom: \cite{DeRosa14, DM91, FM92, Reid+08, Kroupa+03}.
      Adapted from \cite{Thies+15}.  }
    \label{fig:KJ_binfrac}
  \end{figure}

Given that the observed maximum in the stellar luminosity function,
$\Psi_X = \Psi(M_x)$, defined as ${\rm d}N=-\Psi_X\, {\rm d}M_x$, of
all populations has the KTG peak, the remaining bias affecting the
interpretation of $\Psi_X$ into $\xi$ via Eq.~\ref{eq:KJ_xi} is through
unresolved multiple stars.  The observer can only count the number of
main-sequence stars as a function of their luminosity, i.e. construct
$\Psi(M_x)$, but typically 50~per cent of all main-sequence stars with
about $m<{\rm few}\,M_\odot$ are binaries with some being also triple
and quadruple systems \citep{Goodwin+07,Duchene13}. Note that most
stars with mass $<\,$few$\,M_\odot$ should form as binaries
\citep{GK05}. Fig.~\ref{fig:KJ_binfrac} illustrates that, when
including very young stars of all masses, it transpires that the
binary fraction is close to~1 and independent of primary star
mass. Only the old, Galactic field main sequence stars show a trend of
decreasing binary fraction with decreasing primary mass. This is very
well accounted for if all stars form as binaries in embedded clusters
\citep{Thies+15}.
  The birth population of binaries with primaries $<\,$few$\,M_\odot$
  has well defined distribution functions (periods, mass ratios and
  eccentricities), and is well described by randomly sampling the IMF
  for the component masses, leading to a flat mass-ratio distribution
  function \citep{Kroupa95b, Belloni+17}. As emphasised in the latter
  reference, when modelling a stellar population it is essential to
  first draw the desired number of stars from the IMF and then to
  randomly pair the stars from this ensemble to create the binaries,
  because a different procedure would affect the IMF.

  Detailed calculations and modelling of the Milky Way field
  population revealed that the deep star counts (reaching stars to
  distances beyond about 20\,pc which rely on the method of
  photometric parallax to estimate the stellar distances) do not
  resolve most of the multiple systems, or they miss the companions
  due to the flux limit or due to glare from the primary. Taking this
  bias into account, and also the Malmquist and Lutz-Kelker biases,
  leads to a good reproduction of the various empirical determinations
  of $\Psi_V$ \citep{KTG93}. Hereby the novel approach used by
  \cite{KTG93} to arrive at the canonical IMF (for $m<1\,M_\odot$) was
  to employ the observational constraints on the $M_V(m)$ relation,
  the deep and the nearby (i.e. parallax-based) star counts
  simultaneously, to solve for one IMF and for the thickness of the
  Galactic disk. The approach was to assume a binary population in the
  Milky Way disk and to match the deep star counts, which do not
  resolve multiple systems, for the same IMF which also matches the
  nearby star counts in which multiple systems are resolved.  More
  recent constraints on the IMF from MW disk star count data have not
  lead to a revision of these results \citep{Bochanski10}.

Thus, the IMF as derived from the MW disk stellar population can be
written as a two-part power-law form
\begin{equation}
\xi(m) = k_i\, m^{-\alpha_i},
\label{eq:KJ_IMF1}
\end{equation}
where $k_i$ ensure the appropriate normalisation and continuity, and
$\alpha_1=1.3, 0.08 < m/M_\odot \le 0.5$ and
$\alpha_2=2.3, 0.5< m/M_\odot \le 1$ being the Salpeter power-law
index.  A log-normal description is also possible and is
indistinguishable from the above simpler form \citep{Kroupa13}.

\begin{figure}
	\includegraphics[scale=0.9]{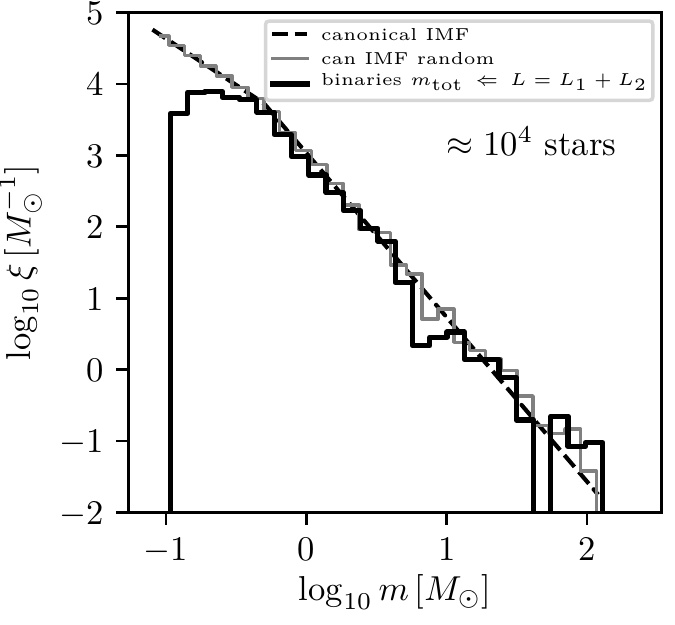}
	\includegraphics[scale=0.9]{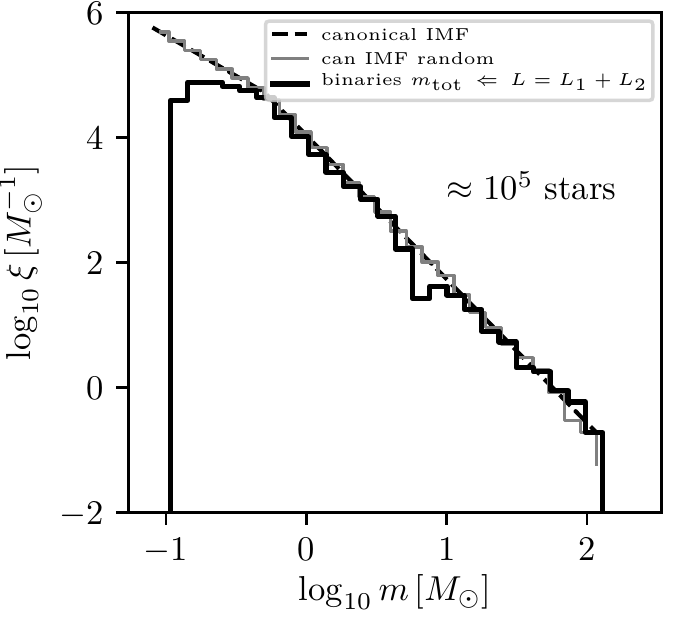}
	\caption[The effect of unresolved binaries to stellar IMF]
	{To demonstrate the effect of unresolved binaries on IMF
          estimates from observations, the system IMF is plotted as
          the thick solid histogram assuming all stars are in
          binaries. The distribution of stars drawn randomly from the
          stellar IMF is shown as the solid gray histogram, and the
          canonical IMF is the dashed line.  Stars are paired into
          binaries as follows: stars more massive than $5\, M_{\odot}$
          are assumed to have a mass-ratio distribution which is
          uniform in the range between 0.1 and 1, as in \cite{Oh15},
          while primaries less massive than $5\,M_\odot$ have
          companions drawn randomly from the IMF leading to a flat
          mass-ratio distribution. Important here is to follow the
          correct procedure as explained in \cite{Oh15} and
          \cite{Oh16}. Note that here the zero-age main sequence
          luminosities of the stars in a binary are added to provide
          the luminosity of the unresolved binary system, which is
          then converted into a mass using the main sequence
          mass--luminosity relation from \cite{Nebojsa2004} and
          \cite{Salaris2006} to define the luminosities of stars of
          mass $m$.  The feature around $5\, M_{\odot}$ is caused by
          the approximate treatment of the bimodal binary
          distribution. The unresolved binaries can significantly
          affect the low mass end of the IMF if not corrected for
          appropriately, but have a negligible effect on the
          massive-star IMF (see also \citealt{WKM09}). }
	\label{fig:KJ_binIMF}
\end{figure}

Taking care of modelling binary systems is an essential part of
arriving at the above result, as neglect of unresolved binary systems
leads to different biases in populations with different dynamical
histories, see Fig.~\ref{fig:KJ_binIMF}. Thus, an observer may readily
derive different "IMFs" in two populations which in fact have the same
IMF but have different binary populations. 

To achieve insights into this problem, the distribution function of
the initial binary population needs to be derived, in equivalence to
the IMF, in order to then be able to calculate the present-day binary
populations for different systems, allowing for the dissolution of the
binaries and the ejection of stars through stellar-dynamical
encounters.  This means mathematically formulating the birth
distributions of orbital periods, mass ratios and eccentricities again
as "hilfskonstrukts" since they are not observable and do not exist at
any instant in time, because, for example, wide binaries may be
dissolved in an embedded cluster before other stars have formed.  This
needs to be tackled to arrive at a full understanding of the Galactic
field population and of individual star clusters. For late-type stars,
much progress has been achieved but requires the inclusion of highly
accurate Nbody modelling currently only possible with the Aarseth
direct Nbody codes and the Giersz-MOCCA Monte Carlo code for massive
clusters \citep{Kroupa95b, Belloni+17}. The consistency check of using
these birth binary distribution functions, evolving typical
birth-CSFEs until they have dissolved to the Galactic field, yields
stellar luminosity functions in very good agreement with the observed
trigonometrical-parallax-based and photometric-parallax-based
luminosity functions and also with the observed field binary population
\citep{Kroupa95MF}.

Fig.~\ref{fig:KJ_binIMF} visualises the effect of binaries on the IMF
by first drawing a list of stars from the IMF. Each star is given a
luminosity and stars weighing $m<5\,M_\odot$ are paired randomly from
this list, while more massive primaries obtain a secondary if the
mass-ratio lies in the range $0.1-1$, both the primary and secondary
being taken from this same list, the stars being then removed from the
list.  This procedure, of only using stars in the list, is important
in order to not change the IMF, for further details see \cite{Oh15}
and \cite{Oh16}. The stellar system ``masses'' are calculated from the
combined luminosity of the binaries, and this yields the ``system
IMF''. It is virtually identical above about $1\,M_\odot$to the
stellar IMF but has a very significant deficit of less massive
stars. Fig.~\ref{fig:KJ_binIMF} demonstrates the maximal error that
can be done when assuming a stellar population has no
binaries. Typical stellar populations have a binary fraction near
50~per cent but the bias at the low mass end is still very large, as
is shown in \cite{Kroupa13} for example. Because only a system IMF can
be deduced from the observations, unless every binary is resolved, the
arrived at IMF would have a turnover in the range $0.3-0.5\,M_\odot$,
which however is not present in reality.

A similar degree of work constraining the birth distributions of
binaries, as is available for $m<\;{\rm few}\;M_\odot$ stars
\citep{Kroupa95b}, is not yet available for massive binaries. The
observational constraints on the present-day binary properties of
massive stars have been quantified well recently \citep{Sana+12,
  MDiS17}, but these need to be evolved backwards through the
dynamical environment of their birth clusters to distill the likely
birth distribution functions. While this is is a mature procedure for
late-type stars \citep{Kroupa95b,Belloni+17}, only the first steps
into this direction have been made by \cite{Oh15} and \cite{Oh16}.
 
Concerning the massive stars, the seminal work of \cite{Massey03}
showed the observationally constrained IMF to be the canonical
invariant Salpeter power-law with $\alpha_3=2.3, m>1\,M_\odot$,
independent of metallicity and density for very young stellar
populations available in the Local Group of galaxies (see
Eq.~\ref{eq:KJ_IMF1} for the low-mass stars). Impressive here is that
many different teams using different telescopes and observing
different clusters and OB associations have shown the IMF to be
essentially invariant with the canonical Salpeter-Massey index
$\alpha_3=2.3, m>1\,M_\odot$ (fig.5 in \citealt{Kroupa02}).  That is,
the shape of the IMF not showing the expected Poisson scatter, and the
$m_{\rm max}-M_{\rm ecl}$ relation having little or no intrinsic
scatter \citep{Weidner13b}, together confirm the IMF to be closer to
an optimal density distribution function (Sec.~\ref{sec:KJ_natIMF}).
Binary and higher-order multiple systems do not affect $\alpha_3$
significantly (\citealt{WKM09} and Fig.~\ref{fig:KJ_binIMF}), although
changes in component masses through binary-stellar-evolution affect
the power-law index to some degree \citep{Schneider+15}, and the
merging of binary components induced in the stellar-dynamically
violent environment of massive embedded clusters lead to the
appearance of super-canonical ($m>150\,M_{\odot}$) stars
\citep{Banerjee12}.

However, given the large internal binding energy of binaries amongst
massive stars, massive stars are ejected preferentially from their
birth clusters. Noteworthy here is that the canonical IMF is perfectly
consistent with the observationally derived IMF in modest M31 star
clusters by \cite{Weisz15}, given that massive stars are ejected from
these (fig.8 in \citealt{Oh16}).  When observed, these massive stars
are missing, leading to a deficit of massive stars in the young
clusters. This has been noted to be the case for the Orion Nebula
Cluster \citep{pflamm-altenburg2006a}. This bias has been corrected
for in the star-burst cluster R136 in the Dor~30 star forming region
in the Large Magellanic Cloud, showing that this
$\approx 10^5\,M_\odot$ very young low-metallicity cluster is likely
to have been born with a top-heavy IMF \citep{Banerjee12b}. This has
been confirmed recently by independent observations
(\citealt{Schneider+18}, see also \citealt{Kalari+18} for another
similar case).

\section{The IMF is a systematically varying function} 
\label{sec:KJ_IMFvar}

From a theoretical point of view \citep{Larson98, AF96, AL96, Dib07}
it has been relatively straight-forward to understand that the IMF
ought to become top-heavy with decreasing metallicity (less capability
for the gas to cool) and increasing density (more coagulations of gas
clumps, stronger converging flows). In regions of high star formation
rate, cosmic rays produced by the high rate of supernovae penetrate
molecular clouds leading to higher ambient temperatures and thus most
probably also to top-heavy IMFs \citep{Papadopoulos11,
  Papadopoulos13}. The IMF may become somewhat bottom-heavy in
high-metallicity clouds and bottom-light at low metallicities for the
same reasons. Essentially, the above theoretical framework largely
rests on star-formation depending on the local Jeans mass and/or being
self-regulated through the various feedback processes that act once
proto-stars appear, whereby photons have a larger interaction cross
section with higher-metallicity gas, while it is modulated by the
gravitational potential which, if deep enough, may cause proto-stellar
cores to merge before individual stars can form in them.  On the other
hand, the theory based on gravo-turbulent molecular clouds
\citep{Padoan+07, HC08, Hopkins13} appears to be challenged by recent
and rather surprising observations according to which stars from in
fine, phase-space coherent fibres and filaments \citep{Hacar+17,
  Hacar+18, Bresnahan+18}, which can only appear when the cloud has
stopped being turbulent on the relevant scales.  A strongly turbulent
cloud can also not lead to a bottom heavy IMF as suggested by
\cite{Chabrier+14}, because the shocks destroy low-mass proto-stellar
cores before they can collapse \citep{Bertelli+16, Liptai+17}. Given
the immensely complex physical processes and boundary conditions acting
during star formation, it is not surprising that it has until now not
been possible to arrive at a convincing theoretical description of the
properties of the IMF such that it is consistent with observed stellar
and brown dwarf populations.

Rather than relying on theoretical arguments, the line of thought
followed here is that the properties of the stellar IMF are
constrained from observed simple stellar populations (containing stars
of the same metallicity and age) in order to extract rules according
to which it changes with conditions. Such an approach allows star
formation theory to be tested and, independently of how successful
star formation theory may be, it ensures that the IMF is consistent
with resolved stellar populations. It allows larger systems such as
galaxies to be described as being composed of many simple
populations. A embedded, open or globular cluster is typically composed
of a simple population.

The above alured-to top-heaviness has been confirmed by direct
star-counts in the 30~Dor region \citep{Schneider+18}.  Other direct
evidence for a top-heavy IMF in a massive low-metallicity star cluster
has also recently been published \citep{Kalari+18}.  An elaborate
analysis of the evolution of Milky Way globular clusters (GCs) and of
ultra compact dwarf galaxies (UCDs) has previously \citep{Dabring09,
  Dabring12, Marks12} allowed a quantification of the dependency of
the shape of the IMF on metallicity and density of the
embedded-cluster forming cloud core,
\begin{equation}
\alpha_{1,2,3}= {\rm fn}({\rm [Fe/H] }, \rho_{\rm ecl}),
\label{eq:KJ_IMF2}
\end{equation}
where [Fe/H] is the iron abundance relative to the Solar value and
$\rho_{\rm ecl}$ is the mass-volume-density (stellar plus gas mass) of
the embedded cluster prior to gas expulsion (eq.~12, 14 and 15 with
0.87 replaced by $-0.87$ in \citealt{Marks12}).  According to these
results the IMF becomes increasingly top-heavy above a
star-formation-rate density of about $0.1\,M_\odot/($yr pc$^3)$ within
the embedded-cluster-forming cloud core on a~pc~scale \citep{Marks12,
  Kroupa13}.  Further evidence for this systematic variation of the
stellar IMF with metallicity and density comes from the mass-to-light
ratia of star clusters in the Andromeda galaxy \citep{Zonoozi16,
  Haghi17}. The high rate of core-collapse supernovae in the central
UCD-type objects in the star-bursting galaxy Arp~220
\citep{Dabring12}, the larger rates of type~Ib and~IIb relative to
'normal' type~II supernovae in Arp~299 \citep{Anderson+11} as well as
the different ratios of supernova types in star-forming (metal-poor)
dwarf relative to (metal-rich) giant galaxies found in the Palomar
Transient Factory \citep{Arcavi+10} all suggest the above systematic
MF variation.

\begin{figure}
    \includegraphics[scale=0.9]{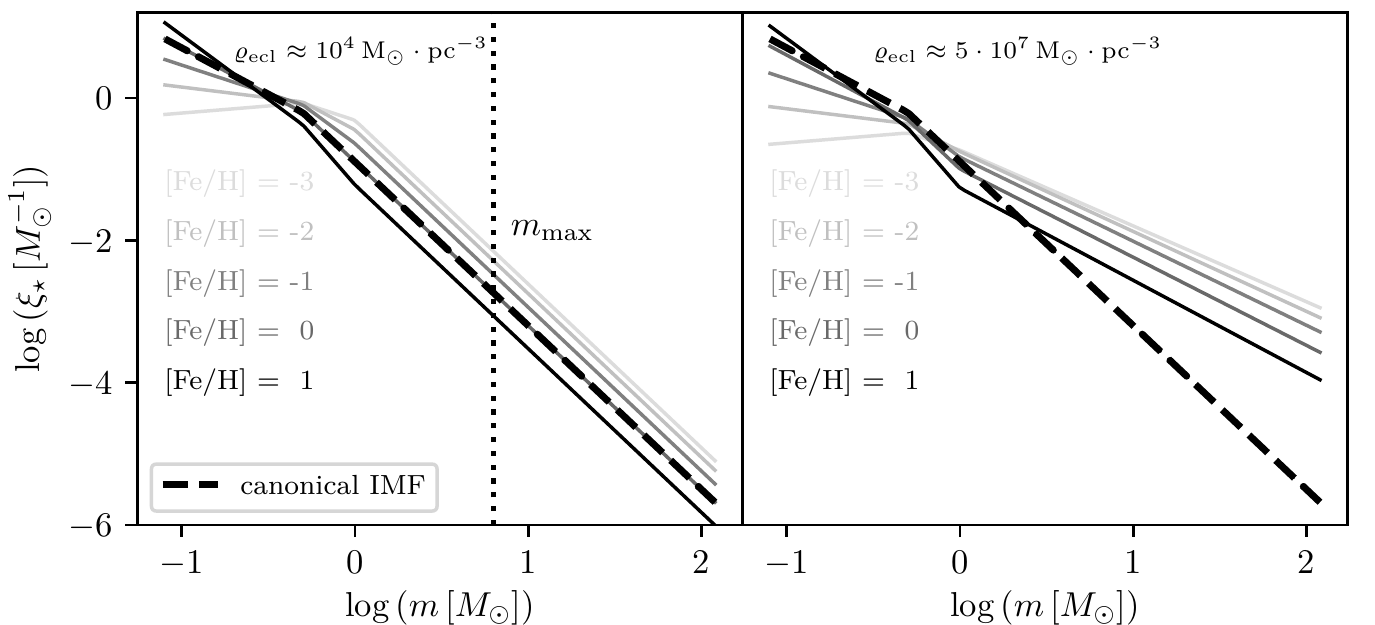}
    \caption[The variable stellar IMF]
    {The variable IMF in comparison to the canonical IMF
      ($\alpha_1=1.3, \alpha_2 = \alpha_3 = 2.3$) shown as the dashed
      line. There is significant observational evidence that the IMF
      is not universal, that is, that it depends on environment, see
      Eq.~\ref{eq:KJ_IMF2}.  Based on \cite{Marks12}, it is shown here
      how the IMF can vary with density and metallicity. In each panel
      the metallicity and density-dependent IMF is shown as a solid
      line of thickness as given by the key in the panel.  Left panel:
      low-density case, corresponding to an embedded star cluster with
      a stellar mass of $M_{\rm ecl}=100\,M_{\odot}$. The vertical
      line demonstrates the expected high mass star cut-off based on
      the $m_{\mathrm{max}}-M_{\mathrm{ecl}}$ relation
      \citep{Weidner13b}, that is, such an embedded cluster would not
      contain stars more massive than $m_{\rm max}$ unless a binary
      merges.  Right panel: High-density case, corresponding a
      $10^8\,M_{\odot}$ star cluster in which
      $m_{\rm max}\approx 150\,M_\odot$. The density,
      $\varrho_{\mathrm{ecl}}$, is given in terms of the mass in gas
      and stars within the embedded cluster, as defined in eq.~7 in
      \cite{YJK17}. All IMFs are normalised to the same stellar
      mass. }
    \label{fig:KJ_varIMF}
  \end{figure}

This variation of the IMF is visualised in Fig.~\ref{fig:KJ_varIMF},
noting that the bottom-lightness at low metallicity may be related to
the bottom-light trend with decreasing metallicity found by
\cite{Gennaro+18} for ultra-faint dwarf galaxies \citep{Jerab18}. The
top-heaviness at low metallicity and high density may well be relevant
for the emergence of multiple populations of stars in globular
clusters \citep{PC06, BJK17}.

  With this we now have, for the first time, a mathematical
  formulation of a systematically varying IMF which is consistent with
  resolved nearby very young and open star clusters, GCs and UCDs and,
  it appears, also whole galaxies \citep{KW03, Kroupa13, Recchi15,
    YJK17, Watts+18, Jerab18}. Elliptical galaxies, which formed with
  star formation rates larger than a few thousand $M_\odot/$yr, will
  have enriched with metals rapidly, as already inferred by
  \cite{Matteucci94} and \cite{GM97}, leading, in their innermost
  super-solar abundance regions, to stellar IMFs which are
  bottom-heavy \citep{Conroy+12}, since the high-metallicity
  individual star-burst embedded clusters forming in the innermost
  region of such systems will have had bottom-heavy IMFs (eq.~12 in
  \citealt{Marks12}).  The application of Eq.~\ref{eq:KJ_IMF2} to
  galaxies, by adding all IMFs in all forming embedded clusters,
  resolves the H$\alpha$ radial cutoff vs the UV-extended
  galactic-disk problem \citep{PAK08}, implies all galaxies,
  independent of mass, to consume their gas on an about~$3\,$Gyr
  timescale, solving the stellar-mass-build-up time problem of dwarf
  galaxies \citep{Pflamm09a} and leads to the observed
  mass--metallicity relation of galaxies \citep{KWK07, Fontanot17}.

An application of Eq.~\ref{eq:KJ_IMF2}  to the early Universe
  suggests that the very young and very massive GC and UCD progenitors
  must have appeared quasar-like with immense luminosities and
  supernova~II rates, opening tests of IMF variation to high-redshift
  studies \citep{Jerab17}.

\bibliographystyle{cambridgeauthordate}
\bibliography{literature}

 \copyrightline{} 
 \printindex
    
\end{document}